\documentclass[aps,pra,twocolumn,amsmath,amssymb,showpacs,floatfix]{revtex4-1}

\usepackage{graphicx,url}

\newcommand{\ket}[1]{\ensuremath{|{#1\rangle}}}

\newcommand{\ketbra}[2]{\ensuremath{|{#1 \rangle}{\langle #2}|}}
\newcommand{\op}[1]{\hat{#1}}

\newcommand{\I}{\text{i}}

\usepackage[colorlinks,citecolor=blue,urlcolor=blue,linkcolor=blue,hyperindex,breaklinks]{hyperref}

\begin{document}

\title{Scheme for the protective measurement of a single photon \\using a tunable quantum Zeno effect}
\author{Maximilian Schlosshauer}
\affiliation{Department of Physics, University of Portland, 5000 North Willamette Boulevard, Portland, Oregon 97203, USA}

\begin{abstract} 
This paper presents a proof-of-principle scheme for the protective measurement of a single photon. In this scheme, the photon is looped arbitrarily many times through an optical stage that implements a weak measurement of a polarization observable followed by a strong measurement protecting the state. The ability of this scheme to realize a large number of such interaction--protection steps means that the uncertainty in the measurement result can be drastically reduced while maintaining a sufficient probability for the photon to survive the measurement. \\[-.1cm]

\noindent Journal reference: \emph{Phys.\ Rev.\ A\ }\textbf{97}, 042104 (2018), DOI: \href{http://dx.doi.org/10.1103/PhysRevA.97.042104}{10.1103/PhysRevA.97.042104}
\end{abstract}


\maketitle

\section{Introduction}

Protective measurement \cite{Aharonov:1993:qa,Aharonov:1993:jm,Aharonov:1996:fp,Dass:1999:az,Vaidman:2009:po,Gao:2014:cu} is a special kind of weak quantum measurement \cite{Aharonov:1988:mz,Duck:1989:uu} that enables the measurement of expectation values of observables on a single system while the disturbance of the system's initial quantum state during the measurement can be made arbitrarily small. Applications of protective measurement include the measurement of the quantum state of a single system \cite{Aharonov:1993:qa,Aharonov:1993:jm,Aharonov:1996:fp,Dass:1999:az,Vaidman:2009:po,Auletta:2014:yy,Diosi:2014:yy,Aharonov:2014:yy,Schlosshauer:2016:uu}, determination of stationary states \cite{Diosi:2014:yy}, investigation of particle trajectories \cite{Aharonov:1996:ii,Aharonov:1999:uu}, translation of ergodicity into the quantum realm \cite{Aharonov:2014:yy}, studies of fundamental issues of quantum measurement \cite{Aharonov:1993:qa,Aharonov:1993:jm,Aharonov:1996:fp,Alter:1997:oo,Dass:1999:az,Gao:2014:cu}, and the complete description of two-state thermal ensembles \cite{Aharonov:2014:yy}. 

Recently, the first experimental realization of a protective measurement has been reported by Piacentini \emph{et al.\ }\cite{Piacentini:2017:oo}, implementing a version of a protective measurement that is based on the quantum Zeno effect \cite{Aharonov:1993:jm,Kwiat:1999:zz,Itano:1990:oo}. In the experiment, a single photon prepared in a polarization state $\ket{\psi}=\cos\theta \ket{H}+\sin\theta\ket{V}$ passes through $N=7$ identical optical stages, each consisting of a birefringent crystal and a polarizer. The birefringent crystal imposes a small polarization-dependent shift in the transverse direction, thus coupling a polarization observable of the system (the polarization degree of freedom the photon) to the apparatus pointer (the spatial mode of the photon). The thickness of the birefringent crystal is chosen such that the beam separation for orthogonal polarizations remains incomplete (weak measurement). After this measurement interaction, a polarizer projects the photon back onto the initial state $\ket{\psi}$, realizing the state protection. After the photon has passed through all $N$ interaction--protection stages, its position is registered by a spatially resolving single-photon detector. 

The experiment demonstrated the shift of transverse photon position by an amount proportional to the expectation value of the measured photon polarization observable, thereby revealing information about an expectation value in the course of a single measurement \cite{Piacentini:2017:oo}. Also, a weak value \cite{Aharonov:1988:mz,Duck:1989:uu} has been obtained from a measurement on a single system, rather than from an ensemble. While the protection procedure requires knowledge of the quantum state, the protective measurement nonetheless offers an important advantage over conventional strong (projective) measurements. Specifically, the Zeno protective measurement typically provides a far better estimate of the expectation value (in the sense of smaller uncertainty in the measurement result) than could be achieved, using comparable resources, from strong measurements on an ensemble of photons \cite{Piacentini:2017:oo}. 

Since the action of the birefringent crystal has changed the photon state, there is a nonzero probability for the photon not to make it past the state-protecting polarizer, leading to photon loss and thereby to an unsuccessful measurement. This is the quantum-Zeno analog of the state disturbance induced by a continuous (non-Zeno) protective measurement \cite{Schlosshauer:2016:uu,Schlosshauer:2015:uu,Schlosshauer:2014:pm,Schlosshauer:2014:tp}. By decreasing the shift of the beam generated by a single birefringent crystal (that is, by decreasing the interaction strength), the probability of the photon reaching the output  after passage through all $N$ stages can be increased. However, this will also decrease the total shift of the photon position at the output, leading to greater uncertainty in the expectation value measured from this shift, especially when the total shift is not significantly larger than the FWHM of the spatial mode of the single photons. To compensate, one may enlarge the number $N$ of interaction--protection stages. While doing so does decrease the photon survival probability, the decrease grows very slowly with $N$, much slower than the decrease in uncertainty; for $N=100$ and a moderately weak measurement, the photon survival probability is still in excess of 50\% \cite{Piacentini:2017:oo}. 

Therefore, in order to optimize the quality of the protective measurement, it is desirable to significantly increase the number $N$ of interaction--protection stages over the $N=7$ stages used in the experiment of Piacentini \emph{et al.\ }\cite{Piacentini:2017:oo}. To enable this increase in $N$, this paper describes a proof-of-principle scheme in which the photon is looped repeatedly through the same interaction--protection stage before it is switched out after an adjustable (and possibly large) number $N$ of iterations (see Ref.~\cite{Kwiat:1999:zz} for a similar approach unrelated to protective measurement). Then, by choosing a birefringent crystal that induces a very small beam shift compared to the beam width and letting the photon traverse many times $N$ through the loop containing the interaction--protection stage, one would in principle be able to realize a high-quality protective measurement. The scheme can be implemented using commonly available optical devices. 

This paper is organized as follows. Section~\ref{sec:prot-meas} briefly reviews the theory of protective measurement applied to the case of photon polarization. Section~\ref{sec:experimental-scheme} describes the scheme for the protective measurement of single photons using a tunable quantum Zeno effect. Section~\ref{sec:noise} remarks on the influence and mitigation of optical noise processes that occur during the measurement.  Section~\ref{sec:discussion} presents a concluding discussion.

\section{\label{sec:prot-meas}Theoretical background}

Consider a photon prepared in the initial quantum state $\ket{\Psi_i}=\ket{\psi} \ket{\phi(x_0)}$, where $\ket{\psi}=\cos\theta \ket{H}+\sin\theta\ket{V}$ represents the polarization state of the photon and the spatial mode $\ket{\phi(x_0)}$ is represented by a Gaussian of width $\sigma$ centered at $x_0$,
\begin{equation}
\ket{\phi(x_0)} \,\dot{=} \,\phi_{x_0}(x) = \left( \frac{1}{2\pi\sigma^2}\right)^{1/4} \exp \left[-\frac{(x-x_0)^2}{4\sigma^2}\right].
\end{equation}
Suppose we let the photon pass through a birefringent material that displaces the horizontally polarized component ($H$) by an amount $+\kappa$ and the vertically polarized component ($V$) by an amount $-\kappa$. This interaction is described by the Hamiltonian
\begin{equation}
\op{H}_\text{int} = \kappa \left( \ketbra{H}{H}-\ketbra{V}{V}\right)\otimes \op{P},
\end{equation}
representing a measurement of the polarization observable $\op{O}=\ketbra{H}{H}-\ketbra{V}{V}$, with $\op{P}$ generating the polarization-dependent shift of the center of the Gaussian wave packet. The measurement strength is quantified by the beam-displacement parameter $\kappa$ relative to the width $\sigma$ of the spatial mode, i.e., by the ratio $\xi = \kappa/\sigma$. For the relevant case of weak measurement ($\xi \lesssim 1$), photon polarization is only incompletely encoded in the spatial degree of freedom. 

After the interaction, the polarization degree of freedom of the photon is projected back onto the initial state $\ket{\psi}=\cos\theta \ket{H}+\sin\theta\ket{V}$, realizing the protection. Assuming the measurement interaction is weak, after $N$ such interaction--protection steps the final photon wave function $\ket{\Psi_f}$ is given by \cite{Aharonov:1993:jm, Dass:1999:az,Gao:2014:cu,Schlosshauer:2015:uu}
\begin{align}\label{eq:vjhgsjh}
\ket{\Psi_f} & \approx \ket{\psi} \exp\left( -\frac{\I}{\hbar} N \kappa  \langle \op{O} \rangle \op{P} \right) \ket{\phi(x_0)} \notag \\ &= \ket{\psi} \ket{\phi(x_0+N\kappa  \langle \op{O} \rangle)}.
\end{align}
Thus, the center of the wave packet is shifted by $N\kappa  \langle \op{O} \rangle$, where $\langle \op{O} \rangle= \cos^2\theta-\sin^2\theta$ is the expectation value of $\op{O}$ in the state $\ket{\psi}$. In the limit $\kappa \propto 1/N$ with $N\rightarrow \infty$ (i.e., an infinitely weak interaction with infinitely many interaction--protection steps), the evolution \eqref{eq:vjhgsjh} becomes exact. This realizes an ideal protective measurement, in which information about the expectation value of $\op{O}$ is encoded in the spatial mode of the photon without disturbing the polarization state of the photon. For finite measurement strengths, one cannot avoid state disturbance \cite{Schlosshauer:2014:pm,Schlosshauer:2015:uu}, here manifesting as photon loss at the protection stage \cite{Piacentini:2017:oo}. Since the wave-packet shift is approximately equal to $N\kappa$, decreasing $\kappa$ requires increasing the number $N$ of interaction--protection steps to maintain appreciable total beam displacement relative to the beam width $\sigma$ (i.e., $N\xi \gg 1$), such that the uncertainty of the measurement result is not unduly increased \cite{Piacentini:2017:oo}. It is this desired increase in $N$ that motivates the scheme described in the next section.

\section{\label{sec:experimental-scheme}Experimental scheme}

\begin{figure*}
\centering \includegraphics[scale=.95]{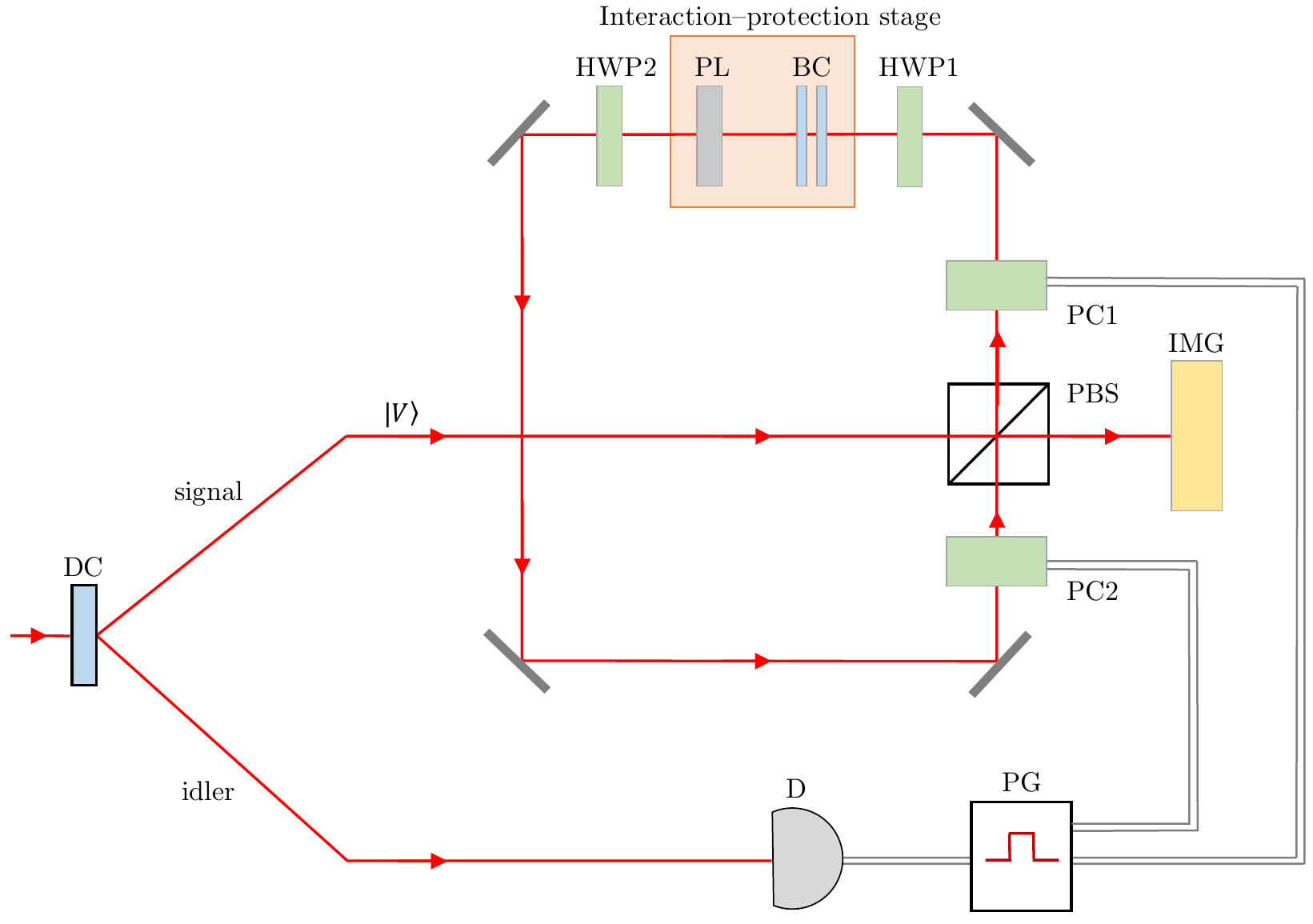}
\caption{\label{fig:scheme}(Color online) Proposed scheme for a photonic implementation of a protective measurement based on the quantum Zeno effect. A vertically polarized signal photon resulting from a down-conversion process (DC) is reflected into the optical loop at the polarizing beam splitter (PBS). Upon leaving the PBS, brief activation of a Pockels cell (PC1) by a pulse generator (PG), which is triggered by the arrival of the idler photon at a detector (D), changes the polarization of the signal photon to horizontal. The signal photon then traverses an interaction--protection stage consisting of two half-wave plates (HWPs), a pair of birefringent crystals (BCs), and a linear polarizer (PL). HWP1 prepares the initial quantum state $\ket{\psi}=\cos\theta \ket{H}+\sin\theta\ket{V}$. BC implements a small polarization-dependent spatial displacement of the photon, weakly coupling polarization and spatial degrees of freedom. PL realizes the state protection by projecting the photon back onto the initial state $\ket{\psi}$. HWP2 rotates the polarization to horizontal, thereby readying the photon for transmission at the PBS for its next round trip through the loop. After $N$ such round trips, a second Pockels cell (PC2) is activated to rotate the polarization to vertical and switch the photon out of the loop to be detected at the spatially resolving photon imager (IMG).}
\end{figure*}
 
Figure~\ref{fig:scheme} shows the proposed proof-of-principle scheme for the protective measurement of a single photon using a tunable quantum Zeno effect. A vertically polarized photon from a photon pair produced by parametric down-conversion is incident on a polarizing beam splitter (PBS) from the left, as shown. The photon is reflected upward at the PBS and describes a round trip through the loop as depicted in the figure. In the absence of further optical devices in the loop, the photon would be reflected at the PBS and exit the loop. To keep the photon in the loop, a Pockels cell (PC1) is briefly activated when the photon first enters the loop, switching the photon's polarization by $90^\circ$ to horizontal  during that first pass such that the photon will be transmitted at the PBS on each subsequent pass (a similar technique has been used for optical storage loops \cite{Pittman:2002:pp}). The activation of the Pockels cell is produced by a pulse generator that is triggered by the arrival of the second (idler) photon at a single-photon detector. The pulse length is chosen such that the Pockels cell is turned off before the photon enters the second cycle through the loop.

A half-wave plate (HWP1) rotates the horizontal polarization of the photon to prepare an arbitrary ``initial'' quantum state $\ket{\psi}=\cos\theta \ket{H}+\sin\theta\ket{V}$. The photon then passes through an interaction--protection stage as used in Ref.~\cite{Piacentini:2017:oo}. The stage consists of a pair of birefringent crystals and a linear polarizer oriented at angle $\theta$ from the horizontal. The first birefringent crystal implements the weak, polarization-dependent beam displacement. Because the birefringence introduces a time and phase delay between the polarization components, a second birefringent crystal is used to compensate for the delay. The polarizer then projects the photon onto the state $\ket{\psi}$, realizing the state protection. Finally, HWP2 rotates the polarization of the photon back to horizontal, thus preparing the photon for transmission at the PBS and its next round trip. 

After a predefined number of round trips have been completed, a second Pockels cell (PC2) is activated to rotate the polarization of the photon to vertical, switching it out of the loop. The photon and its position are then detected by the photon imager placed at the output of the PBS (Piacentini \emph{et al.\ }used a $32\times 32$ array of silicon single-photon avalanche diodes \cite{Piacentini:2017:oo}). Thus, activation of PC1 and PC2 mark the beginning and end, respectively, of the protective measurement, with the delay time between activation of PC1 and PC2 determining the desired number of cycles. The zero position of the photon at the imager (i.e., the position without beam displacement that results in the absence of the weak measurement) can be defined by initially removing the polarizer and birefringent crystals from the setup and not activating either of the Pockels cells, such that an incident photon is switched out of the loop after a single round trip.

\section{\label{sec:noise}Influence and mitigation of noise}

The above discussion has neglected the loss and noise processes associated with the devices in the optical loop. In practice, cross-talk in the polarizing beam splitter means that not all horizontally polarized photons will be transmitted, and inaccuracies in setting the polarizer and wave-plate angles, together with polarization changes induced by the mirrors, may cause deviations from the desired polarization states inside the optical loop. For example, if the photon entering the polarizing beam splitter after a pass through the loop has acquired a vertically polarized component, or if the beam splitter incorrectly reflects a horizontally polarized photon, then the photon might prematurely exit the loop before having completed the desired number of round trips. In this case, the beam displacement of such a photon would be smaller than the amount expected based on the full number of round trips, leading to an underestimation of the expectation value encoded in the beam position. 

Given that the loop will in general be traversed a large number of times, such noise effects may accumulate. However, the fact that the Zeno scheme prescribes that the photon must be returned to its initial polarization state after each passage through the loop suggests a way of mitigating at least some of the noise effects in a manner that avoids their accumulation. For example, by placing an additional linear polarizer, oriented to pass horizontal polarization, immediately before the second Pockels cell (PC2 in Fig.~\ref{fig:scheme}), one can actively reinforce proper photon polarization after each pass through the loop (albeit at the expense of an increase in the probability of photon loss). Additionally, one could mitigate the influence of premature exits by timing the detection at the photon imager such that the photon is counted only if it has completed the desired number of round trips. Thus, a photon that has prematurely exited will simply be discarded. Proper detector timing can be suitably defined by the start trigger provided by the coincident detection of the idler photon, together with the expected travel time of the photon in the loop.  

Photon loss inside the loop due to absorption at optical devices is another limiting factor. However, this problem is not unique to this scheme, but occurs in any Zeno-type setup in which the photon passes many times through an array of identical optical devices. While the absolute size of such losses depends on the quality of the optical devices and the accuracy of their alignment, it is equally important to consider how such practical losses compare to the fundamental probability of photon loss intrinsic to the measurement scheme itself. As mentioned above, the intrinsic loss probability is on the order of 50\% for $N=100$ passes and a moderately weak measurement \cite{Piacentini:2017:oo}, so it is not unreasonable to expect that in many situations such intrinsic losses will substantially outweigh the optical losses arising from imperfections. It also bears noting that in those situations where the main goal of the experiment is a faithful measurement result for those photons that do complete the optical loop, loss processes do not necessarily need to be considered problematic.

While the photon polarization is reset after each pass through the loop, the spatial displacement of the photon is accumulated over many passes. As a consequence, imperfections in the displacement incurred during each pass will also accumulate. It will therefore be of paramount importance to ensure that the beam inside the optical loop is level, with the beam displacement occurring only in a horizontal plane and in a well-defined relation to the placement of the birefringent crystals.

\section{\label{sec:discussion}Discussion}

The scheme proposed here aims to provide a flexible implementation of a high-quality photonic protective measurement. Its key component is a timed optical loop that allows the photon to pass through an arbitrary and easily adjusted number $N$ of interaction--protection stages. In such an experiment, one may use a birefringent crystal with very small beam displacement relative to the beam width, and then increase the number $N$ of loop cycles until a sizable shift of the photon position is seen on the photon imager. In this way, the expectation value may be measured with low uncertainty while ensuring a large photon survival probability. It is expected that a successful implementation of the present scheme would substantially improve the performance of the setup described in Ref.~\cite{Piacentini:2017:oo}. 

In practice, the typical loss and noise processes associated with any optical device may dictate a reasonable upper limit for $N$ if we are to maintain an acceptably large photon survival probability. Measures have been mentioned for mitigating noise processes inside the optical loop that avoid the problem of accumulation of errors during repeated passes. It would be interesting to see how an experimental realization of the scheme will perform in the vicinity of $N=100$, the number theoretically considered in Ref.~\cite{Piacentini:2017:oo}. 

It is noted that the quantum Zeno technique has also been employed in the realization of so-called interaction-free measurements \cite{Elitzur:1993:oo,Kwiat:1995:za,Kwiat:1999:zz}. The basic idea of an interaction-free measurement is to infer the presence of a quantum object without interacting with it. In the original, non-Zeno scheme of such a measurement \cite{Elitzur:1993:oo}, the presence of an opaque object in one arm of an interferometer is inferred from the detection of a probe photon at the previously dark output of the interferometer, without the photon having interacted with the object. In the quantum Zeno version of an interaction-free measurement \cite{Kwiat:1995:za,Kwiat:1999:zz}, an initially horizontally polarized probe photon undergoes $N$ repeated small polarization rotations $\Delta \theta = \pi/2N$. After each rotation, the photon passes through a polarization interferometer that has an object placed in the vertically polarized arm. Transmission of the photon through the interferometer therefore amounts to projecting the photon onto the initial, horizontal polarization state. By increasing $N$, the probability for the photon to be absorbed by the object can be made arbitrarily small. Thus, with the object present, the polarization of the photon after $N$ passes will be horizontal, while in the absence of the object the final polarization will have been rotated to vertical. Measurement of this final polarization therefore reveals information about the presence of the object, even though the photon has not interacted with the object. While the Zeno-type sequence of projection steps in this scheme is similar to the repeated projections in the protective measurement described in this article (and in Ref.~\cite{Piacentini:2017:oo}), the interaction-free measurement and the protective measurement are otherwise rather different. The interaction-free measurement ascertains the presence of an object by making inferences from the absence of an interaction; indeed, the projection can be viewed as a consequence of the non-interaction. The protective measurement, by contrast, includes an interaction between system and probe on each pass (in the present case, photon polarization is coupled to the spatial degree of freedom), and the purpose of the projection is to disentangle the system and probe after each interaction step to protect the initial photon state. Moreover, rather than measuring presence, the protective measurement measures the expectation value of an arbitrary observable of the system. 

\begin{acknowledgments}
This research was supported by the M. J. Murdock Charitable Trust.
\end{acknowledgments}


%

\end{document}